\documentclass[aps,prl,twocolumn,superscriptaddress]{revtex4-1}

\usepackage{amssymb}
\usepackage{graphicx}
\usepackage{amsmath}
\usepackage{braket}
\usepackage{hyperref}
\usepackage{bbold}

\begin{document}

\title{Matrix Product State Representation without explicit local Hilbert Space Truncation with Applications to the Sub-Ohmic Spin-Boson Model}

\author{Max F. Frenzel}
\affiliation{Blackett Laboratory, Imperial College London, Prince Consort Road, London SW7 2BW, UK}
\author{Martin B. Plenio}
\affiliation{Blackett Laboratory, Imperial College London, Prince Consort Road, London SW7 2BW, UK}
\affiliation{Institut f\"ur Theoretische Physik, Albert-Einstein-Allee 11, Universit\"at Ulm, D-89069 Ulm, Germany}

\date{\today}

\begin{abstract}
We present an alternative to the conventional matrix product state representation, which allows us to avoid the explicit local Hilbert space truncation many numerical methods employ. Utilising chain mappings corresponding to linear and logarithmic discretizations of the spin-boson model onto a semi-infinite chain, we apply the new method to the sub-ohmic SBM. We are able to reproduce many well-established features of the quantum phase transition, such as the critical exponent $\frac{1}{2}$ predicted by mean-field theory. Via extrapolation of finite-chain results, we are able to determine the infinite-chain critical couplings $\alpha_c$ at which the transition occurs and, in general, study the behaviour of the system well into the localised phase.
\end{abstract}


\maketitle

\section{Introduction\label{Introduction}}
The spin-boson model (SBM) describes a single two-level system (TLS), a spin, coupled to environmental degrees of freedom represented by a continuous bath of bosonic field modes. It is one of the most important models for studying the general effects arising when a quantum system is coupled to an environment \cite{Leggett1987}. In the sub-ohmic version, it possesses a mean-field like quantum phase transition between a localised and a delocalised phase. This quantum phase transition has been the subject of extensive numerical and analytical investigations \cite{Chin2011a,Chin2010,Vojta2005,Alvermann2009,Chin2006,Silbey1984,Winter2009,Guo2012}. Yet, many numerical approaches face challenges near and above the transition to the localised phase. This is due to the rapidly rising number of field excitations in the localised phase, which imply that the quantum states of the field modes span an increasingly large subspace of their full Hilbert space. Most numerical methods are however based on local Hilbert space truncation and are hence discarding increasing amounts of vital information.\\
In this paper, we describe a variation on the matrix product state (MPS) representation that avoids the explicit local Hilbert space truncation using a soft cut-off instead. We demonstrate the usefulness of this approach by applying it to study the properties of the second-order magnetic quantum phase transition of the sub-ohmic SBM and their comparison to an analytical approach based on a variational ansatz \cite{Chin2011a}.

\section{MPS Method} 

Generally, the compound quantum state of an $M$-level system (whose states are labelled $k = 1,...,M$) and an environment of $N$ bosonic modes (with states labelled $i_1,...,i_N \in \mathbb{N}^0$) can be described by
\begin{equation} \label{eq:state}
\ket{\psi} = \sum_{k, i_1,..., i_N} c_{k, i_1,...,i_N}\ket{k, i_1,...,i_N}.
\end{equation}
Alternatively, it is possible to express the coefficients $c_{k, i_1,...,i_N}$ as products of $\chi \times \chi$ matrices, where the M-level system as well as each mode have a unique set of matrices associated with them
\begin{align}\label{eq:MPS_original}
\ket{\psi} = & \sum_{k, i_1,..., i_N} tr[S^{(k)}\prod_{m=1}^N B_m^{(i_m)}] \prod_{m=1}^N(b_m^{\dagger})^{i_m}\ket{k, 0} \nonumber \\
= & \sum_{k, i_1,..., i_N} tr[S^{(k)}\prod_{m=1}^N B_m^{(i_m)}\sqrt{i_m!}] \ket{k, i_1,...,i_N},
\end{align}
where $\ket{0}$ represents the vacuum state of all modes and $b_m^{\dagger}$ is the creation operator of the $m$-th mode.\\
This representation of the state is known as a matrix product state (MPS) \cite{Ostlund1995,Rommer1997,Perez-Garcia2007,Fannes1992}. In this form the state of the $M$-level system is represented by the $M$ matrices $S^{(1)},...,S^{(M)}$ and the $m$-th field mode is represented by a semi-infinite set of matrices $\{B_m^{(i_m)}\}$. In this infinite number of matrices lies the problem of the traditional MPS ansatz. Numerical calculations are limited to a finite set of matrices, hence the local Hilbert space associated with each mode has to be truncated to a finite size by limiting the local dimension. Under certain conditions, such as high mean excitation numbers, this truncation can lead to substantial errors in numerical calculations.\\
Here we consider an alternative MPS-type representation that retains the ability to represent correlations between subsystems and avoids the hard truncation of the local Hilbert space dimensions. This is achieved by reducing the number of matrices per mode to a single matrix $X_m$, which is defined such that
\begin{equation} \label{eq:X}
B_k^{(i_m)} = \frac{X_m^{i_m}}{i_m!},
\end{equation}
for $0\leq i_m < \infty$, i.e. the infinite set $\{B_m^{(i_m)}\}$ is now formed from powers of a single matrix $X_m$, reducing the total number of matrices required to fully describe the state to $M+N$ \footnote{The fact that this method always associates the identity matrix with the ground state did not appear to be a problem in the simulations presented here, but it might be an area for further investigation.}. The additional factor of $(i_m!)^{-1}$ is chosen to simplify later calculations. In addition, instead of directly restricting the bosonic Hilbert space to a finite set of low occupation sates, it introduces a "soft cut-off" in the number of allowed bosons, giving lower occupational states a higher weight than states with large boson number, and vanishing weight in the limit $i_m \rightarrow \infty$. It should be noted that this representation shares some semblance to a MPS representation in a variable coherent state basis which becomes transparent when diagonalizing the matrix $X$. As we admit arbitrary forms for $X$ that are obtained in the optimization part of our algorithm, our approach does not require a specific choice of basis but determines the optimal choice automatically.\\ 
Substituting into \eqref{eq:MPS_original}, the MPS can be written as
\begin{equation} \label{eq:MPS}
\ket{\psi} = \sum_{k=1}^M\sum_{i_1,..., i_N = 0}^{\infty} tr[S^{(k)}\prod_{m=1}^N \frac{X_m^{i_m}}{\sqrt{i_m!}}] \ket{k, i_1,...,i_N}.
\end{equation}
This form not only enables us to avoid a direct truncation of the bosonic Hilbert space, but also continues to allow for straightforward determination of the normalization as well as expectation values. Introducing the new notation
\begin{equation} \label{eq:Lambda}
\Gamma_a^b \equiv \prod_{m=a}^b e^{\bar{X}_m \otimes X_m}
\end{equation}
and
\begin{equation} \label{eq:S}
\Xi \equiv \sum_{k=1}^M \bar{S}^{(k)} \otimes S^{(k)},
\end{equation}
we find for the norm 
\begin{equation} \label{eq:norm}
\braket{\psi | \psi} = tr[\Xi \Gamma_1^N]. 
\end{equation}
The exponentiation of the $X_m$ matrices in combination with the deliberate choice of the $(i_m!)^{-1}$ factor results in the exponential functions appearing in eq. \eqref{eq:Lambda}, which are straightforward to evaluate numerically.\\ 
As an example of an expectation value, we find for the population of the $k$-th mode
\begin{equation} \label{eq:expectation}
\braket{\psi | b_k^{\dagger} b_k | \psi} = tr[\Xi \Gamma_1^k (\bar{X}_k \otimes X_k) \Gamma_{k+1}^N]. 
\end{equation}
In general, we find that all quantities of interest are simple traces over products of $\chi^2 \times \chi^2$ matrices.\\
A ground state MPS in the form of eq. \eqref{eq:MPS} can be found for an arbitrary Hamiltonian $H$ by starting with a state $\ket{\psi}$ formed of randomly chosen $S$ and $X$ matrices and then minimising the energy
\begin{equation} \label{eq:Energy}
E = \frac{\braket{\psi | H | \psi}}{\braket{\psi | \psi}}. 
\end{equation}
with respect to $\ket{\psi}$, i.e. finding the $S$ and $X$ matrices which minimise eq. \eqref{eq:Energy}.\\
The approach we are taking here is heuristic. It is motivated by the desire to reduce the number of free parameters in the description of the ansatz wavefunction, while retaining the essential features of the physical wavefunction. Indeed, if we consider a single system only, then we find that every state admits a representation as in eq. \eqref{eq:MPS} for sufficiently large (possibly infinite) matrices. We expect, but have not proven, that this remains true for multipartite states too. Correlations between subsystems can be described increasingly well by using growing matrix dimensions, following the philosophy of matrix product states. Besides this, the demonstrated computational efficiency and the lack of a hard cut-off are additional points in favour of this approach.\\
To further demonstrate the viability and usefulness of this ansatz we now apply this method to analyze the ground state properties of the sub-ohmic spin-boson model.

\section{Spin Boson Model\label{SBM}}
The Hamiltonian of the (unbiased) spin-boson model (SBM) is given by ($\hbar=1$) 
\begin{equation} \label{eq:Hamiltonian_SBM}
H_{SB} = -\frac{1}{2}\Delta\sigma_x + \frac{1}{2}\sigma_z\sum_l g_l(a_l + a_l^{\dagger}) + \sum_l\omega_l a_l^{\dagger} a_l,
\end{equation}
where $\sigma_i$ are the usual Pauli matrices describing a two-level system (TLS) with tunnelling amplitude $\Delta$. $a_l$ and $a_l^{\dagger}$ are the bosonic annihilation and creation operators of the environment, which consists of bath modes with frequency $\omega_l$. The key quantity in the description of the system-environment interaction is the spectral function $J(\omega)=\pi\sum_lg_l^2\delta(\omega-\omega_l)$. Here we consider a spectral function of the form 
\begin{equation} \label{eq:spectrum}
J(\omega) = 2\pi\alpha\omega_c^{1-s}\omega^s\Theta(\omega_c-\omega),
\end{equation}
as given in \cite{Bulla2005}, where $\omega_c$ is the maximum cut-off frequency of the spectrum and $\Theta(\omega) = 1- \Theta(-\omega) = 1$ for $\omega>0$. In the following we focus exclusively on the sub-ohmic case for which $0<s<1$, in particular on the case $s<0.5$, which we will compare to existing results in the literature.\\
In \cite{Chin2010,Prior2010} it was shown that a Hamiltonian of the form eq. \eqref{eq:Hamiltonian_SBM} can be mapped exactly onto a semi-infinite chain of bosonic modes that experience nearest neighbour interaction only, with the system only coupling to the first chain site. The transformed Hamiltonian can be written as
\begin{eqnarray} \label{eq:Hamiltonian_chain}
H = &-&\frac{1}{2}\Delta\sigma_x + c_0\sigma_z(b_0 + b_0^{\dagger})\nonumber \\
&+&\sum_{n=1}^{\infty}\omega_n b_n^{\dagger}b_n + t_n(b_{n+1}^{\dagger}b_n + b_n^{\dagger}b_{n+1}),
\end{eqnarray}
where the coupling strength between the TLS and the first site is given by
\begin{equation} \label{eq:c}
c_0 = \sqrt{\frac{\alpha}{2(s+1)}}\omega_c
\end{equation}
and the local energies and tunnelling amplitudes of the sites are
\begin{equation} \label{eq:omega}
\omega_n = \frac{\omega_c}{2}\Bigl(1 + \frac{s^2}{(s+2n)(2+s+2n)}\Bigr)
\end{equation}
and
\begin{equation} \label{eq:t}
t_n = \frac{\omega_c(1+n)(1+s+n)}{s+2+2n)(3+s+2n)}\sqrt{\frac{3+s+2n}{1+s+2n}}
\end{equation}
respectively. This mapping brings several advantages, which include the analytical forms for the parameters of the resulting chain model given above, an intuitive picture of how irreversibility emerges \cite{Chin2011} and the ready applicability of the MPS method \cite{Chin2010,Chin2011a}.\\
Utilizing the MPS eq. \eqref{eq:MPS}, we can now find the ground state of the spin-boson system by minimizing eq. \eqref{eq:Energy}. Specifically, we have to find the $S$ and $X$ which minimize 
\begin{equation}\label{eq:E_SB}
E =  \braket{\psi | H | \psi} \equiv E_{loc} + E_{int} + E_{chain},
\end{equation}
with the TLS's local energy
\begin{equation} \label{eq:t}
E_{loc} = -\frac{1}{2}\Delta\braket{\sigma_x}, 
\end{equation}
the system-chain interaction energy
\begin{equation} \label{eq:t}
E_{int} = c_0\braket{\sigma_z(b_0 + b_0^{\dagger})},
\end{equation}
and the chain energy
\begin{equation} \label{eq:t}
E_{chain} = \sum_{n=1}^{\infty}\omega_n \braket{b_n^{\dagger}b_n} + t_n(\braket{b_{n+1}^{\dagger}b_n} + \braket{b_n^{\dagger}b_{n+1}}),
\end{equation}
subject to the constraint $\braket{\psi | \psi} = 1$.
In terms of the MPS description we obtain, after truncating the chain length to N sites, the total energy
\begin{widetext}
\begin{align}\label{eq:E_tot}
E = &-\frac{\Delta}{2} tr\Bigl[\Bigl(\bar{S}^{(1)}\otimes S^{(2)} + \bar{S}^{(2)}\otimes S^{(1)} \Bigr)\Gamma_1^N\Bigr] + 
c_0 tr\Bigl[\Bigl(\bar{S}^{(1)}\otimes S^{(1)} - \bar{S}^{(2)}\otimes S^{(2)}\Bigr)\Bigl(\mathbb{1}\otimes X_1 + \bar{X}_1 \otimes \mathbb{1} \Bigr)\Gamma_1^N\Bigr] \nonumber \\
&+ \sum_{n=1}^{N-1}\Bigl\{tr\Bigl[\Xi\Gamma_1^n\bigl(\omega_n(\bar{X}_n\otimes X_n) + t_n(\bar{X}_{n+1}\otimes X_n + \bar{X}_n\otimes X_{n+1})\bigr)\Gamma_{n+1}^N\Bigr]\Bigr\}
+ \omega_N tr\Bigl[\Xi\Gamma_1^N\bigl(\bar{X}_N\otimes X_N)\Bigr].
\end{align}
\end{widetext}
This minimization can be carried out numerically to yield the full ground state MPS of the TLS and the N-site chain. In the following we will present some results for the ground state properties of the sub-ohmic SBM. The minimizations in this work were carried out using MATLAB's \textit{fminunc} function.

\section{Results}

\begin{figure}[t]
\includegraphics[width = \columnwidth]{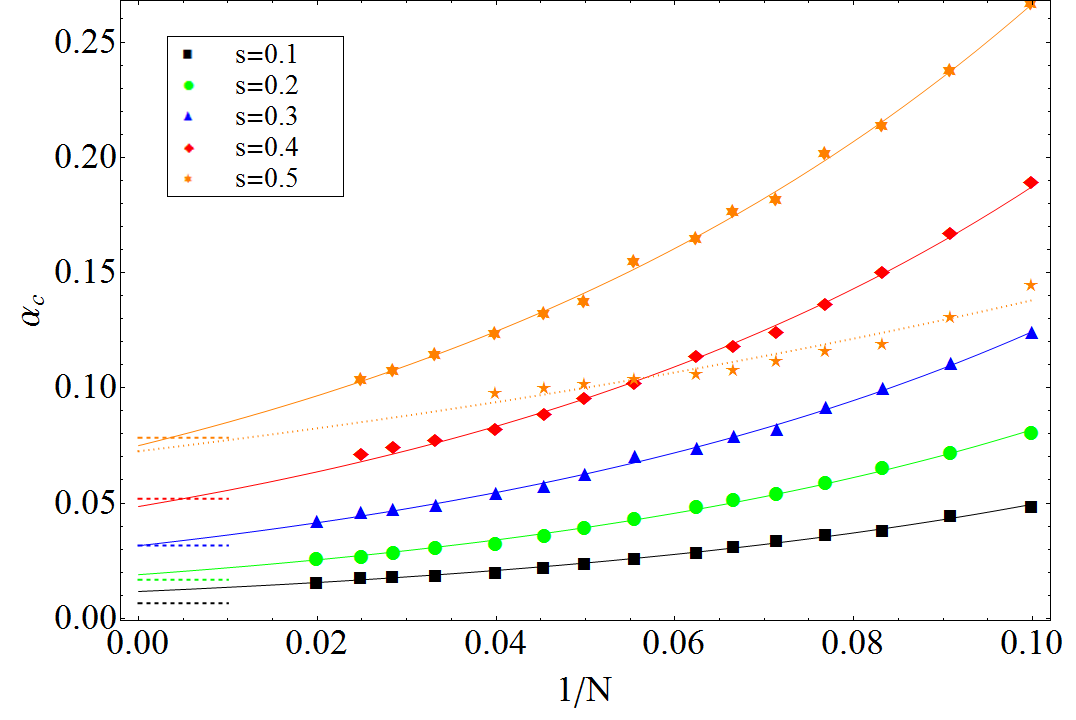}
\caption{Critical coupling $\alpha^{(3)}_c$ as a function of the inverse chain length $1/N$ for $\Delta = 0.1$ and $\omega_c = 1$. The fitted functions are of the form $\alpha^{(3)}_c(N) = ae^{b/N}$, where $a$ is the extrapolated limiting value for $\alpha^{(3)}_c$ as ${N\rightarrow\infty}$. The dotted line for $s=0.5$ corresponds to logarithmic discretisation ($\Lambda = 1.5$) and the short dashed lines are the critical values $\tilde{\alpha}_c$ predicted by the polaron ansatz in \cite{Chin2011a}. The discrepancy between our results and the predicted values hint at a likely inaccuracy of the polaron ansatz, particularly at low $s$.\label{Extrapolation}}
\end{figure}
The sub-ohmic SBM is believed to possess a mean-field-like continuous phase transition in the magnetisation for ${0<s<0.5}$ at a critical coupling strength $\alpha_c$ between system and environment. For small coupling strengths $\alpha < \alpha_c$  the TLS is in a delocalised phase, having no net magnetisation. Above the critical point $\alpha > \alpha_c$ the environment induces a spontaneous magnetisation on the TLS, which then exhibits a doubly degenerate localised phase.\\
In the delocalised phase the mean site population is expected to diverge along the chain. This has so far made it difficult to study the system close to and above the phase transition with great accuracy employing numerical methods that rely on Hilbert space truncation, such as numerical renormalisation group (NRG) \cite{Vojta2005} and density matrix renormalisation group (DMRG) methods \cite{White1992,Schollwock2005}. We propose that this truncation and many of the associated problems can be avoided using the MPS of form eq. \eqref{eq:MPS} and the minimization eq. \eqref{eq:E_tot}.\\
However, in line with the other approaches mentioned above, one still has to perform a different kind of truncation to make the numerical simulation feasible, namely truncation of the chain length $N$, which effectively amounts to an infrared cut-off that is neglecting low frequency components of the environmental bath.\\ 
Fig. \ref{Extrapolation} shows the critical coupling strengths $\alpha^{(\chi)}_c$, defined as the value of $\alpha$ at which the TLS develops a non-zero magnetisation, plotted against the inverse chain length $1/N$ for matrices of dimension $\chi=3$ and several values of $s$, which we determined using our ansatz. A good fit to the data was found for an ansatz of the form 
\begin{equation}\label{eq:fit}
\alpha_c^{(\chi)}(N) = ae^{b/N},
\end{equation}
which we then fitted for each $s$, enabling us to extract an extrapolated value for $\alpha_c^{(\chi)}$ in the limit $N\rightarrow\infty$. Table \ref{alpha_c} shows the results of the extrapolation. In \cite{Chin2011a} a variational ansatz was used to predict some of the properties of the sub-ohmic SBM ground state. For the critical coupling strength they predict a value
\begin{equation}\label{eq:alpha_c}
\tilde{\alpha}_c =  \frac{\sin(\pi s)e^{-s/2}}{2\pi(1-s)}\Bigl(\frac{\Delta}{\omega_c}\Bigr)^{1-s}.
\end{equation}
These predicted values are indicated in Fig. \ref{Extrapolation} by dashed lines and are also listed in Table \ref{alpha_c} along with the fractional deviation between our extrapolation results and the predicted values. We find that the predictions agree reasonably well with our result for large $s$ but show significant deviations at smaller $s$. This suggests that the mean-field type ansatz for the environment holds well for larger $s$ but fails for decreasing $s$, possibly as a result of the increasing correlations in the environment.
\begin{table}[b]
\caption{Critical coupling strengths $\alpha^{(3)}_c$ for various values of $s$, together with the respective critical values $\tilde{\alpha}_c$ derived from the polaron ansatz in \cite{Chin2011a} and the fractional deviation.\label{alpha_c}}
\begin{ruledtabular}
\begin{tabular}{| c | c | c | c |}
$s$ & $\alpha^{(3)}_c$ & $\tilde{\alpha}_c$ & $(\alpha^{(3)}_c - \tilde{\alpha}_c)/\tilde{\alpha}_c$\\ \hline
0.1 & 0.0117$\pm$0.0002 & 0.0065 & 0.800 \\ \hline
0.2 & 0.0189$\pm$0.0003 & 0.0168 & 0.125 \\ \hline
0.3 & 0.0315$\pm$0.0004 & 0.0316 & -0.003 \\ \hline
0.4 & 0.0485$\pm$0.0007& 0.0519 & -0.066\\ \hline
0.5 & 0.0749$\pm$0.0009 & 0.0784 & -0.045
\end{tabular}
\end{ruledtabular}
\end{table}
\begin{figure}[t]
\includegraphics[width = \columnwidth]{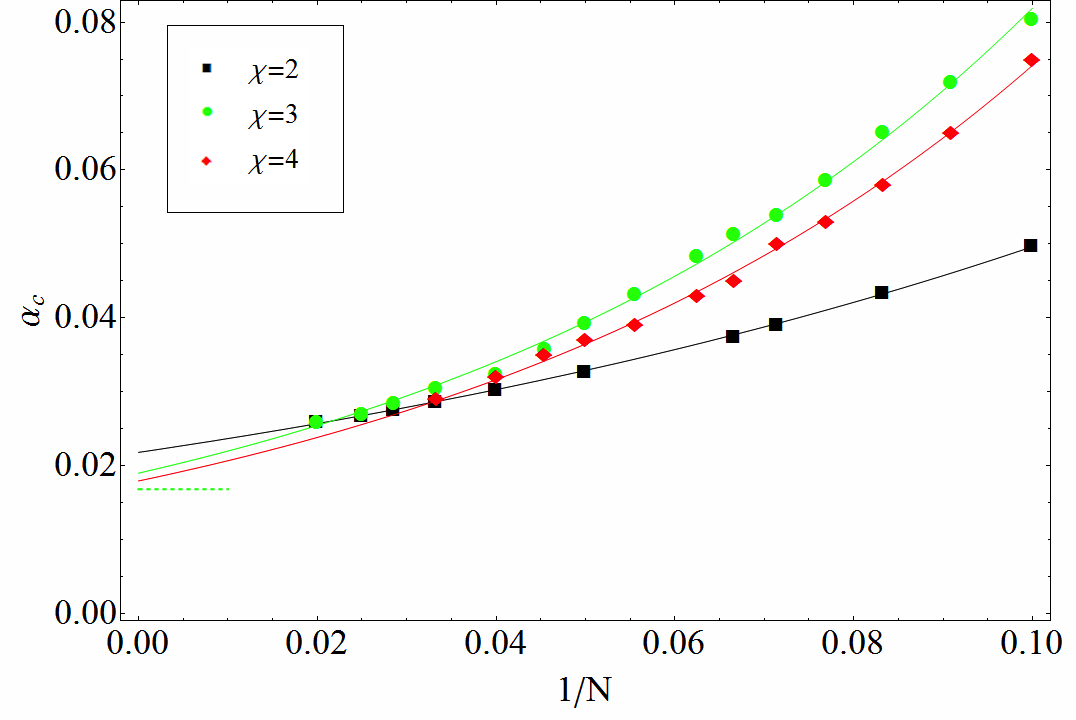}
\caption{Critical coupling $\alpha^{(\chi)}_c$ as a function of the inverse chain length $1/N$ for $s=0.2$, $\Delta = 0.1$ and $\omega_c = 1$ for different matrix dimensions $\chi$. The fitting is of the same form as in Fig. \ref{Extrapolation}. For $\chi=4$ our extrapolated value is $\alpha^{(4)}_c = 0.0179\pm0.0005$ which is in good agreement with $\alpha_c = 0.0175\pm0.0002$ found via Monte Carlo methods in \cite{Winter2009}. \label{ExtrapolationS02}}
\end{figure}\\
In Fig. \ref{ExtrapolationS02} we show the behaviour of the critical coupling strength for different matrix dimensions $\chi$ in the case of $s=0.2$. Even for scalars, $\chi=1$, the system still exhibits a phase transition, but the qualitative behaviour is quite different from $\chi\ge2$. For all $\chi\ge2$ the qualitative features appear to be the same, the only difference being the location of the phase transition as obtained from finite scaling. Via our extrapolation ansatz we find for $s=0.2$
\begin{eqnarray} \label{eq:crit}
\alpha^{(2)}_c = 0.0218\pm0.0001,\nonumber \\
\alpha^{(3)}_c = 0.0189\pm0.0003,\\
\alpha^{(4)}_c = 0.0179\pm0.0005.\nonumber 
\end{eqnarray}
The growing error with increasing $\chi$ is due to the fact that we used the same computation time for all matrix sizes, and hence employed less stringent convergence criteria for higher $\chi$, leading to larger uncertainties on the individual data points. Our value for $\alpha^{(4)}_c$ is in excellent agreement with the critical coupling $\alpha_c = 0.0175\pm0.0002$ found in \cite{Winter2009} using Quantum Monte Carlo simulations for the same set of parameters, showing that, even for very moderate matrix dimensions $\chi$, our method agrees with previous studies. In the following analysis we use $\chi=2$ and $\chi=3$ (to speed up simulations) since in the rest of this article we are mainly interested in the qualitative features of the system, which as mentioned above show no significant deviations for all $\chi\ge2$ in the mean-field regime $0<s\leq0.5$.\\
Being able to find an MPS representation for the ground state, we were also able to analyse the general properties of the state in both the delocalised and the localised phase. Fig. \ref{sigmaz} shows the magnetisation ${M=|\braket{\sigma_z}|}$ of the TLS for some representative values of $s$. This and the following results were obtained with a chain length $N=50$. For $s\leq0.3$ we used $\chi=3$, whereas the results for higher $s$ were obtained using $\chi=2$ matrices due to slower convergence just above the phase transition. Fig. \ref{sigmaz} clearly shows the two phases, separated by a second-order transition. In the delocalised phase $\alpha<\alpha_c$ the order parameter $M$ is zero. Above the critical coupling strength $\alpha_c$ the TLS obtains a finite magnetisation with a tendency to full localisation $M=1$ as $\alpha$ grows large. This localised phase is two-fold degenerate with ${M=\pm\braket{\sigma_z}}$ both being solutions. 
\begin{figure}[t]
\includegraphics[width = \columnwidth]{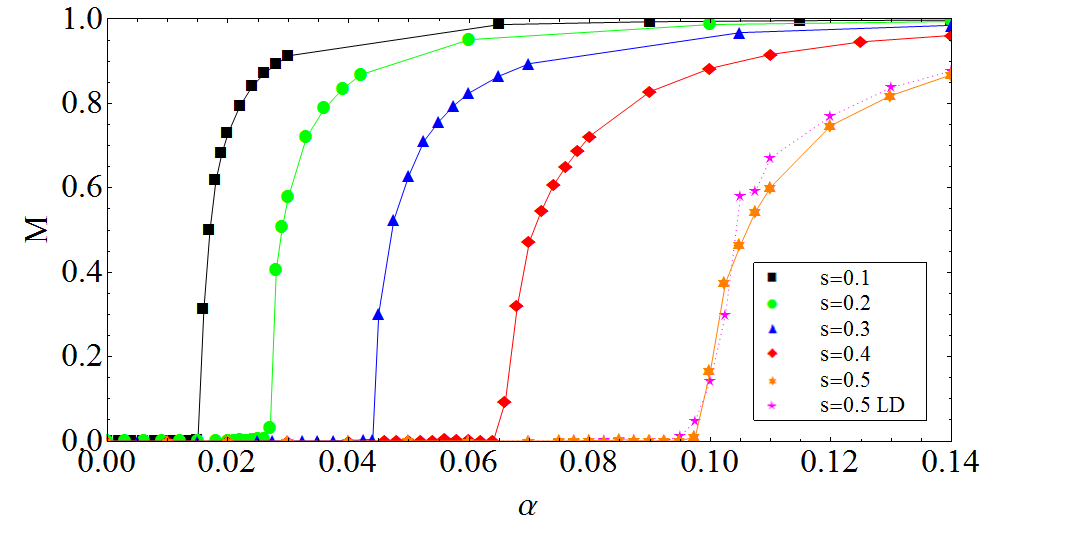}
\caption{Magnetisation $M=|\braket{\sigma_z}|$ as a function of $\alpha$ for $N=50$, $\chi=3$, $\Delta = 0.1$ and $\omega_c = 1$. The second order discontinuity in the order parameter $M$ at $\alpha_c$ marks the phase transition. The magenta points correspond to $s=0.5$ with logarithmic chain discretisation. Data points are joined for better visibilty. Whereas in \cite{Chin2011a} similar plots could only be found through an analytical ansatz, the new method now allows us to obtain numerical results.\label{sigmaz}}
\end{figure}\\
Mean-field theory predicts a second-order magnetic transition at $\alpha_c$ with 
\begin{equation}\label{eq:meanfield}
M \propto |\alpha_c-\alpha|^{1/2}.
\end{equation}
Our simulations do indeed reproduce the correct critical mean-field exponent $\frac{1}{2}$ well, as can be seen in Fig. \ref{Magnetisation}, where we have plotted the magnetisation for $\alpha>\alpha_c$ on a log-log-plot. The mean-field result is indicated in the figure by the solid straight line. To show that the method is also valid in the non-mean-field regime of the sub-ohmic SBM, $0.5<s<1$, we consider the specific case of $s=0.75$, using matrices of dimension $\chi=3$. The results for this case are shown in the inset of Fig. \ref{Magnetisation}. From Fig. 2 in \cite{Guo2012} we expect to find an exponent of approximately 0.25. Our result predicts a scaling according to an exponent of 0.286, which is in reasonable agreement with \cite{Guo2012}, and certainly shows a clear deviation from the mean-field result. This suggests that our method is not limited to the mean-field regime, but can also be applied in other cases.
\begin{figure}[t]
\includegraphics[width = \columnwidth]{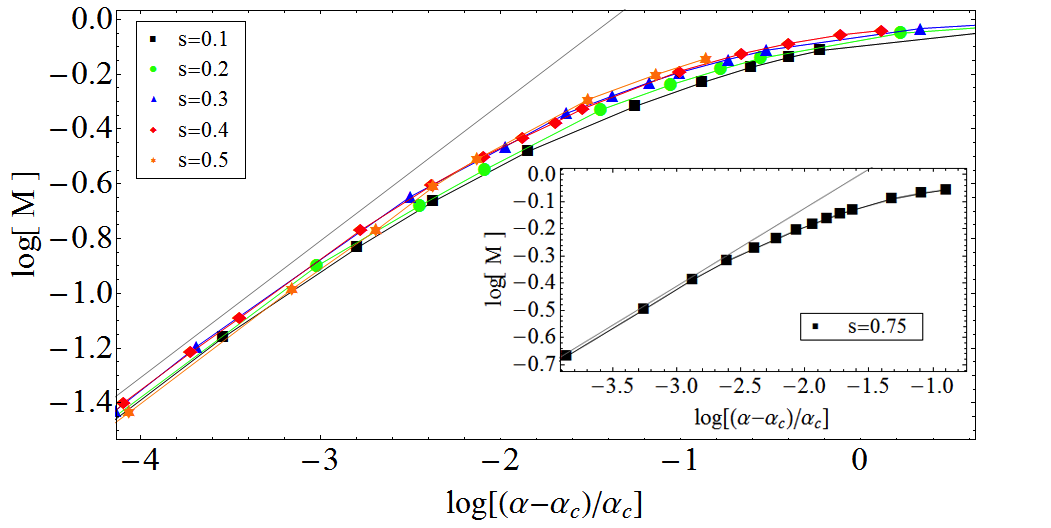}
\caption{Log-Log plot of Magnetisation $M$ as a function of $(\alpha-\alpha_c) / \alpha_c$ for $\alpha>\alpha_c$, $N=50$, $\Delta = 0.1$ and $\omega_c = 1$. The solid straight line represents the expected mean-field exponent $1/2$ just above the critical coupling $\alpha_c$ for $0<s\leq0.5$. The inset shows a similar plot for the case of $s=0.75$ (with $\chi=3$), where we expect a deviation of the exponent from the mean-field value. In this plot the solid straight line represents the exponent 0.286 our result converges to, which appears to be in reasonable agreement with the results presented in \cite{Guo2012}. This suggests that the method presented in this article is also applicable outside the mean-field regime.\label{Magnetisation}}
\end{figure}\\
In \cite{Chin2011a} a variational ansatz was used to predict the amount of entanglement in the TLS, defined as the von Neumann entropy of its reduced density matrix. Our numerical results are presented in Fig. \ref{Entanglement}. Despite the slight deviations in the values for $\alpha_c$, which most likely arise as a combination of the finite chain length we considered as well as the inherent differences between the two approaches (c.f. Table \ref{alpha_c}), our results are in excellent qualitative agreement with the analytical predictions \cite{Chin2011a}. In the delocalised phase entanglement increases. At the critical coupling $\alpha_c$ the entanglement exhibits a cusp and then decays rather rapidly in the localised phase due to the system evolving into a product state. 
\begin{figure}[b]
\includegraphics[width = \columnwidth]{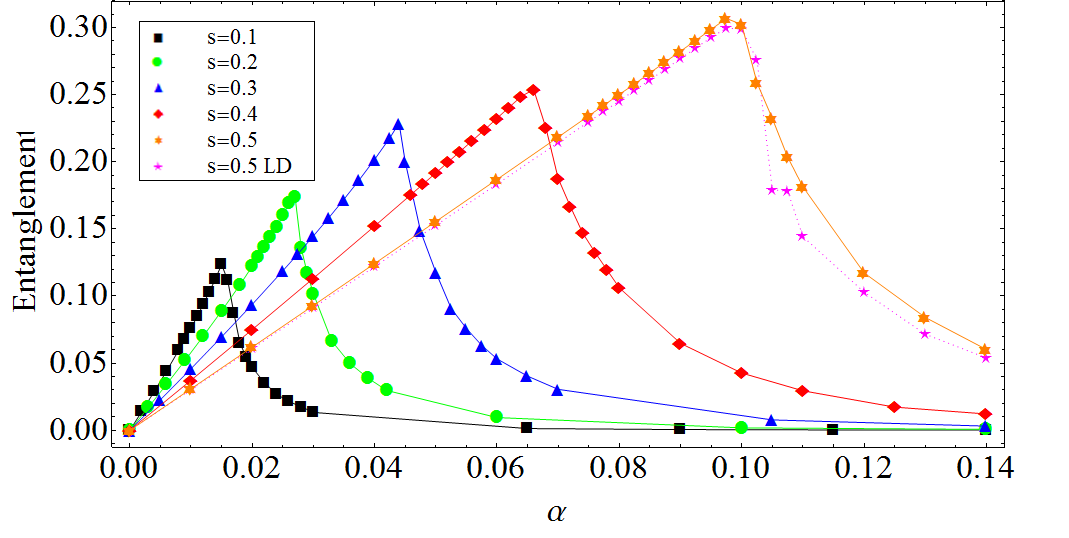}
\caption{Entanglement (von Neumann entropy) between the TLS and the environment for $N=50$, $\Delta = 0.1$ and $\omega_c = 1$. The maxima coincide with the phase transition at $\alpha_c$. The magenta points correspond to $s=0.5$ with logarithmic chain discretisation. \label{Entanglement}}
\end{figure}\\
In addition to the entanglement of the TLS, we also looked at the entanglement of the individual sites in the chain. As a representative example, the results for $s=0.3$ are shown in Fig. \ref{Ent_c} for the first ten chain sites. We find that only the first few sites carry significant amounts of entanglement and that the general behaviour with changing $\alpha$ closely resembles the entanglement properties of the TLS in Fig.  \ref{Entanglement}. A substantial spread of entanglement along the chain is only observable very close to the phase transition.
\begin{figure}[t]
\includegraphics[width = \columnwidth]{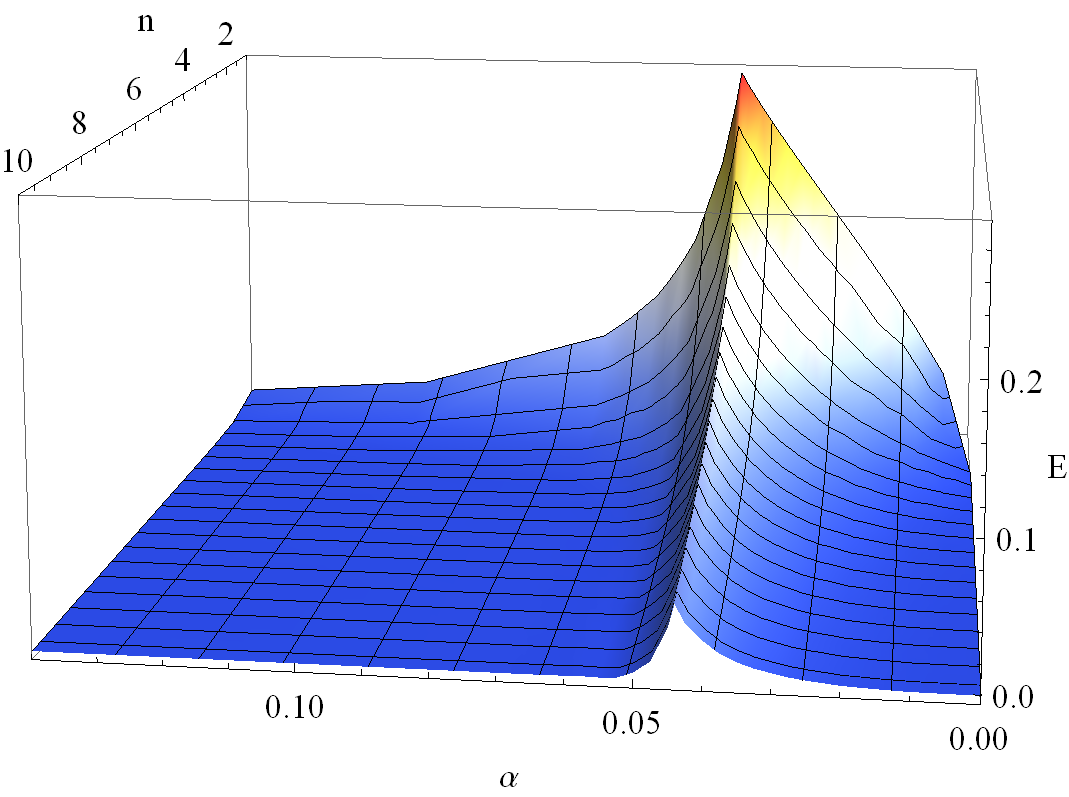}
\caption{Entanglement $E=-tr[\rho_n \log\rho_n]$ of the individual chain sites as a function of $\alpha$ for the first $10$ sites for $s=0.3$ using $\chi=3$. The entanglement of site $n$ is here defined as the von Neumann entropy of the reduced density matrix $\rho_n$ of site $n$. Only the first few sites show significant amounts of entanglement. This shows that the analytical ansatz in \cite{Chin2011a} does have a sound basis in this regime but fails to be accurate near the critical point and perhaps for other choices of $s$ when entanglement is larger.\label{Ent_c}}
\end{figure}\\
Another observable of interest is the coherence $\braket{\sigma_x}$ which is shown in Fig. \ref{sigmax} as a function of $\alpha$. The results are again in excellent qualitative agreement with the results from the variational approach \cite{Chin2011a}. The coherence is continuously decreasing, with a faster decay above the transition, at which we observe a cusp.
\begin{figure}[b]
\includegraphics[width = \columnwidth]{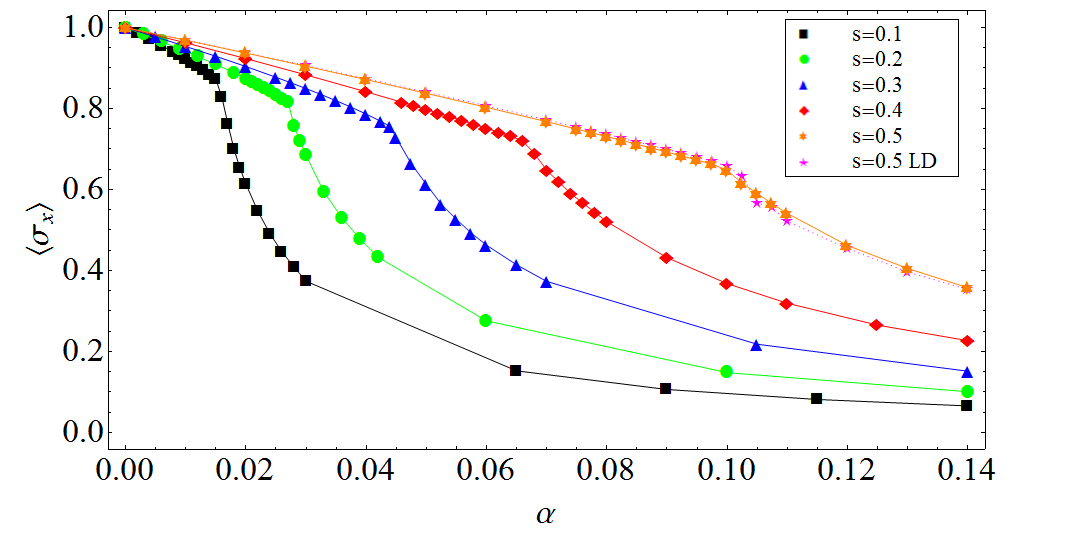}
\caption{Expectation value $\braket{\sigma_x}$ as a function of $\alpha$ for $N=50$, $\Delta = 0.1$ and $\omega_c = 1$. A cusp at $\alpha_c$ marks the phase transition. The magenta points correspond to $s=0.5$ with logarithmic chain discretisation. \label{sigmax}}
\end{figure}\\
As mentioned above, the main reason other numerical approaches have failed to return accurate results near and above the critical coupling is that, whereas in the delocalised phase the mean occupation $\braket{b_n^{\dagger}b_n}$ of chain site $n$ rapidly decreases along the chain, it rises considerably in the localised phase. In fact, \cite{Chin2011a} predicts that  $\braket{b_n^{\dagger}b_n}$ diverges along the chain for $\alpha>\alpha_c$. In Fig. \ref{occupancy} we plot $\braket{b_n^{\dagger}b_n}$ as a function of $\alpha$ and $n$. Below the transition we find that indeed the average population of the sites decreases with $n$. At the transition we observe a sudden increase in $\braket{b_n^{\dagger}b_n}$ and the maximum begins to shift away from the first site, further along the chain. This rapid rise in the occupancy shows why methods such as DMRG, which rely on Hilbert space truncation, are challenged in this regime, since the information about the system is spread over an increasing number of basis states, only a finite number of which are retained by these methods.\\
An alternative method to the linear chain mapping we have thus far considered, is provided by logarithmic discretisation of the spectrum \cite{Bulla2008,Chin2010,Bulla2005,Vojta2005}, which is extensively used in NRG. It does not linearly subdivide the bath spectral function, but instead splits it in intervals $[\Lambda^{-(n+1)},\Lambda^{-n}]$, where $\Lambda>1$ is the discretisation parameter and $n\in \mathbb{N}^0$. This new Hamiltonian has again the same form eq. \eqref{eq:Hamiltonian_chain}, but with different site frequencies and transition amplitudes
\begin{equation} \label{eq:omega2}
\omega_n = \zeta_s(A_n + C_n)
\end{equation}
and
\begin{equation} \label{eq:t2}
t_n = -\zeta_s\frac{N_{n+1}}{N_n}A_n
\end{equation}
respectively, where $\zeta_s$, $A_n$, $C_n$ and $N_n$ are given in \cite{Chin2010}. To be applicable for numerical methods it is again necessary to truncate the resulting chain Hamiltonian at a finite number of sites $n=N$.
\begin{figure}[t]
\includegraphics[width = \columnwidth]{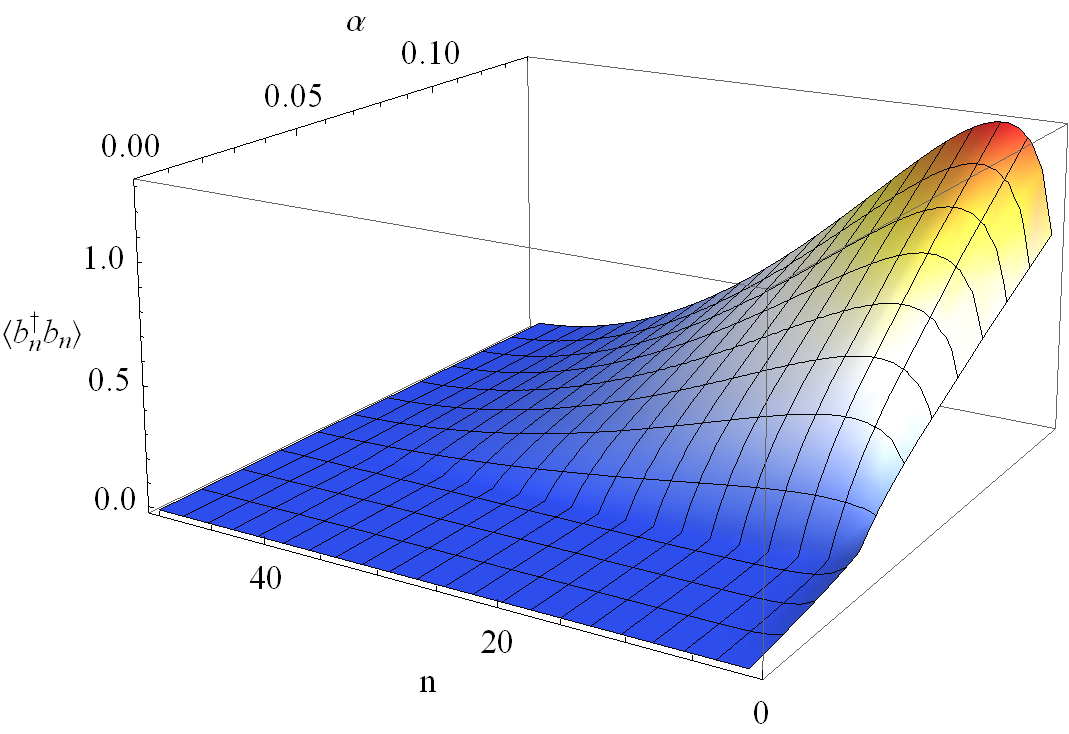}
\caption{Expectation value $\braket{b_n^{\dagger}b_n}$ as a function of $\alpha$ for each chain site (labelled by n) for $s=0.3$ using $\chi=3$. Above the phase transition the mean population of the sites quickly grows, rendering numerical methods such as DMRG which are based on Hilbert space truncation inaccurate in this regime. \label{occupancy}}
\end{figure}\\
A challenge for many numerical methods with the logarithmic discretisation is the fact that the mean occupation of the chain sites is on average considerably larger than on the linearly discretised chain. This quickly leads to the breakdown of these methods. However, as eq. \eqref{eq:MPS} avoids any direct truncation, we were again able to establish numerical results for the sub-ohmic SBM ground state. The dotted line in Fig. \ref{Extrapolation} shows the same extrapolation for $s=0.5$ as carried out for the linear discretisation, now using the logarithmically discretised Hamiltonian with discretisation parameter $\Lambda = 1.5$. Despite the missing data for large $N$, we see that the critical coupling converges to a similar value as found before, $\alpha_c^{LD} = 0.0725\pm0.0029$, with a fractional deviation $|\alpha_c - \alpha_c^{LD}|/\alpha_c^{LD} = 0.033$, where the superscript $LD$ refers to logarithmic discretisation. We also find that with the same chain length $N$, the logarithmically discretised Hamiltonian results in a value for $\alpha_c$ that is generally lower and hence closer to the limiting value for $N\rightarrow\infty$ found via extrapolation than in the linearly discretised case. However, for larger values of $N$, the simulation takes considerably longer to converge near the phase transition due to the comparatively larger Hilbert space that is populated in this scheme. Hence, using the same computational time as for the linear discretisation, we did not acquire reliable data points for $N>25$. The dotted magenta lines for $s=0.5$ in Figures \ref{sigmaz}, \ref{Entanglement} and \ref{sigmax} also show results using the logarithmic discretisation with $N=50$, using exactly the same computational time as the results for linear discretisation. The results are almost identical to those obtained via linear discretisation, except just above the transition at around $0.09\lesssim\alpha\lesssim0.11$ where the convergence issues come into play.
 
\section{Conclusion}
By modifying the traditional matrix product state representation, we were able to avoid the explicit local Hilbert space truncation that leads to the failure of many numerical methods in a regime of high field mode excitation. Instead we introduce a soft cut-off which gives higher weight to lower population numbers, but does not directly truncate the Hilbert space at any dimension. Using this modified state representation combined with a method of energy minimisation, we were able to give a detailed study of the ground state properties of the sub-ohmic SBM. Our findings are in good agreement with previous numerical and analytical results, but extend these to new regimes of the spin-boson model, particularly the region close to and above the phase transition. This regime poses considerable challenges to numerical investigations at present, since methods such as DMRG fail to produce reliable results due to the rapid increase of the Hilbert space dimension of the environmental modes. In addition, our method allowed us to give an analysis of the chain properties such as mode excitation and entanglement for specific sites along the chain. It also has the advantage of being comparatively easy to implement numerically. Hence the method provides a promising new tool to investigate the localised phase of the SBM near and above the transition and to test the current analytical results such as the mean-field type approaches in this regime. The results are particularly remarkable considering the brute-force approach (an inbuilt MATLAB function) used for the minimisation. We believe that further study of the method will provide more specialised techniques, thus speeding up convergence and allowing for efficient simulations with larger matrix dimension $\chi$. Some first applications to other models such as coupled harmonic systems \cite{Plenio2004} also show promising results but require further study to confirm a general applicability of the method.

\section{Acknowledgments}
We acknowledge discussion with Alex W. Chin. This work was supported by the Alexander von Humboldt Foundation.

\bibliography{SBM.bib}

\end{document}